\documentclass[%
 reprint,
superscriptaddress,
 amsmath,amssymb,
 aps,
]{revtex4-1}

\usepackage{graphicx}% Include figure files
\usepackage{dcolumn}% Align table columns on decimal point
\usepackage{bm}% bold math
\usepackage{color}
\begin{document}

\preprint{APS/123-QED}

\title{Can a polarization scrambler really depolarize light?}
%\thanks{A footnote to the article title}%

\author{G. P. Tempor\~ao}
\email{temporao@cetuc.puc-rio.br}
\affiliation{%
 Center for Telecommunication Studies, Pontifical Catholic University of Rio de Janeiro, Rua Marqu\^es de S\~ao Vicente, 225 - G\'avea - Rio de Janeiro - Brazil} 

\author{J. P. von der Weid}
\affiliation{%
 Center for Telecommunication Studies, Pontifical Catholic University of Rio de Janeiro, Rua Marqu\^es de S\~ao Vicente, 225 - G\'avea - Rio de Janeiro - Brazil}

%\collaboration{CLEO Collaboration}%\noaffiliation

\date{\today}% It is always \today, today,
             %  but any date may be explicitly specified

\begin{abstract}
According to quantum theory, two ensembles of quantum systems that are described by the same density operator are indistinguishable. For example, unpolarized light can be obtained either by an incoherent mixture of two orthogonal pure states or by tracing out a photon from a maximally polarization-entangled photon pair. In both cases, one is unable to guess with probability greater than 50\% the outcome of any polarization measurement, but the reasons are conceptually different: whereas the first case is a matter of classical ignorance (the photons were prepared in a definite but unknown way), the second one is of the quantum ignorance kind - if one cannot access the other degrees of freedom of the quantum state, there is no information that could be used to predict a measurement result. We use these concepts to discuss the quantum-physical interpretation of partially polarized light from the point of view of quantum communication and whether a polarization scrambler can really depolarize light. A novel definition of the degree of polarization of a single photon which does not depend on classical ignorance is also provided.

\begin{description}
%\item[Usage]
%Secondary publications and information retrieval purposes.
\item[PACS numbers]
03.67.Dd, 03.67.Hk
%May be entered using the \verb+\pacs{#1}+ command.
%\item[Structure]
%You may use the \texttt{description} environment to structure your abstract;
%use the optional argument of the \verb+\item+ command to give the category of each item. 
\end{description}
\end{abstract}

% PACS, the Physics and Astronomy
                             % Classification Scheme.
%\keywords{Suggested keywords}%Use showkeys class option if keyword
                              %display desired
\maketitle

%\tableofcontents

\section{\label{sec:intro}Introduction}
A \textit{polarization scrambler} is a device that acts as a time-varying unitary transformation on the polarization state of light. They have been used since long ago in long haul optical communications and fiber-optical metrology in order to eliminate polarization sensitivity \cite{Agrawal}, and are commercially available from many manufacturers. In spite of its simple definition, however, the effect of a polarization scrambler over the degree of polarization (DOP) of light has been a subject fraught with controversy. The literature has many examples of schemes that employ scramblers to depolarize light, i.e., to reduce its DOP (possibly to zero, rendering a light beam completely unpolarized) \cite{Heismann, Lize, Karpinski}, and some manufacturers often refer to "depolarizing by polarization scrambling" in the products application notes \cite{Laser2000}. However, most of these schemes do not work as claimed. 

It has already been shown that polarization scramblers generally reduce the \textit{mean} value of the DOP, but not its instantaneous value, defined within the coherence time of the light beam \cite{Matthieu}. However, as all commercial polarimeters measure the mean DOP, the polarization scrambler's effect may become indistinguishable from real depolarization. This happens whenever the polarimeter's detector integration time is greater than the time scale of the fluctuations caused by the scrambler. One possible solution to overcome the detector's response time was also provided by \cite{Matthieu}: by means of a coherent measurement. Given that the DOP is a collective property of a light beam, if one takes any two photons from the same mode, they must have the same polarization state whenever the beam is polarized. This can be verified by projecting the photon pair on the singlet state: two identical photons in the same pure state will always result in zero probability of projection.

The solution above utilizes a quantum measurement for determining the DOP of a classical light beam. But what about single photons? If a nonclassical light source generates one polarized single photon at a time and each photon experiences a different, random polarization rotation by a scrambler, there is no way to distinguish this incoherent mixture of polarized photons from a collection of real unpolarized photons, i.e., photons that have no definite polarization state. The method of projection over the singlet state does not work in this case: as the source produces isolated photons, there is no means to measure a two-photon property. Moreover, even if there were two photons in the same temporal mode, the coherent measurement could give a false result if the photons were in any linear combination of the other three Bell states, as the probability of projection would also be zero. As a matter of fact, the collective measurement approach fails because the DOP from the quantum-physical point of view must be seen as an \textit{individual} rather than a \textit{collective} property. Even though it's impossible to measure the DOP of a single copy of a photon, its polarization state may be well defined; for example, it could have been prepared by passing through a polarizer. As strange and unnatural as it may look, polarization properties of single photons have been extensively used in the last decades in quantum communication applications, such as quantum key distribution \cite{Gisin_RMP}.

Previous works have already pointed out many flaws in the direct adaptation of the classical definition of the DOP to quantum states of light \cite{Bjork,Iskhakov}. Whereas the classical definition can still be appropriate for ensembles of photons, its meaning can be misleading for single photons. In this work, we use the concepts of classical and quantum ignorance to propose a new definition for the DOP of a single photon. We show that the DOP is closely related to the amount of quantum ignorance in a polarization state measurement where we have freedom to measure any observable. In particular, we compare two different quantum communication scenarios where Alice sends unpolarized light (in the classical sense) to Bob, who must guess which of his detectors will fire. Whenever Bob cannot access other degrees of freedom of the photons sent by Alice, he is bound by quantum ignorance.

This work is organized as follows: in section II, the concepts of classical and quantum ignorance are reviewed and discussed. In section III, the boundary between these concepts is illustrated by a quantum communication scenario where Bob can perform arbitrary measurements on the states sent by Alice, whereas section IV discusses the effect of polarization scramblers on single photons. Section V introduces the new proposal for the DOP and section VI draws the conclusions. 

\section{\label{sec:2}Classical vs. Quantum ignorance}
Ever since its inception more than one hundred years ago, quantum physics has awakened an ever-going debate on the interpretation of its bizarre and counterintuitive - but ultimately correct - predictions. Many interpretations of quantum theory seek a "deeper reality" below its pragmatic foundations, particularly in which concerns the probabilistic nature of the universe. As it has turned out, it is difficult to accept a reality where uncertainty and probability play a major role. 

In order to understand the difference between the probabilities that are intrinsic to the quantum-physical description of reality and the probabilities that we are used to experience in our daily lives, we can resort to the concepts of classical ignorance and quantum ignorance \cite{Herbert}. For example, consider a coin toss game where Alice flips a coin, writes down the outcome and sends it by mail to Bob, who must guess the outcome before opening the envelope. Assuming the coin toss was fair, Bob clearly has a 50\% chance of winning the game. Now consider a different guessing game where Alice prepares a single photon and makes it pass through a 50/50 beamsplitter (BS). Each output mode of the BS is connected to an optical fiber, and both are connected to Bob's office. Bob places a single-photon detector in the output of each one of the fibers. Similarly to the first scenario, Bob must guess which of the two detectors will fire before the measurement is performed. Again, he will have 50\% chance of winning.

From a purely probabilistic point of view, the two games are equivalent as Bob has the same odds of correctly guessing the result \cite{Comment1}. From a conceptual point of view, however, the situations are quite different. In the coin tossing game, a probability smaller than one is a consequence of Bob's subjective ignorance on that particular result, i.e., there was a definite result, but it was known only to Alice. This is what we call "classical" ignorance: Bob could have guessed correctly if he had received classical information on the result (e.g., Alice could have told him beforehand over the phone). However, in the single photon experiment, even though Alice has prepared the quantum state that has been sent to Bob, she has no information on which way the photon will go after the BS. According to quantum theory, the most complete description that we can have on the photon is given by its quantum state, which would be of the form $|\phi\rangle = |a\rangle + e^{i\theta}|b\rangle$ (up to a normalization constant), where $a$ and $b$ are the BS output modes. Therefore, the information on which path has been followed by the photon simply does not exist: it is unknown not only to Bob, but to every possible observer in the Universe. This is what we call quantum ignorance.

It is tempting to say that classical and quantum ignorance can be quantified by Shannon's and von Neumann's entropies, respectively \cite{Nielsen}. However, this association is not as straightforward as it seems: pure states have zero von Neumann entropy, but they can lead to maximum quantum ignorance, depending on the measurement being performed. Moreover, even though classical ignorance and quantum ignorance are very distinct concepts, there are some situations where it seems that all ignorance is ultimately classical. Both points are discussed in more detail in the next section, in the context of polarization state measurements.

\section{\label{sec:3}Classical and quantum ignorance in incoherent mixtures of pure states}
\begin{figure}[htb]
\centerline{\includegraphics[width=8cm]{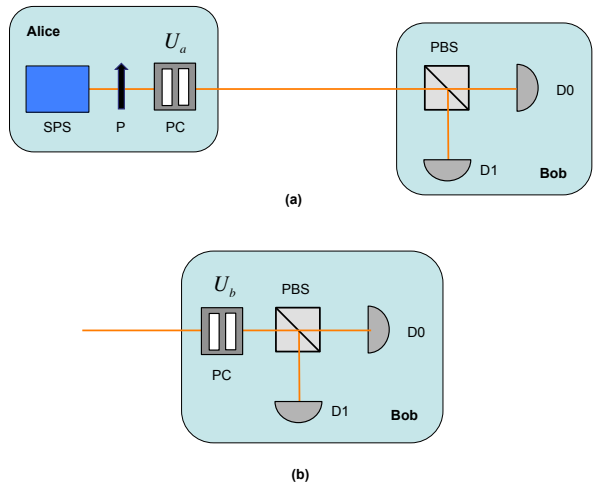}}
\caption{Quantum communication scenario that illustrates the concepts of classical and quantum ignorance in a polarization state measurement. (a) Bob cannot change his measurement basis, which may give rise to quantum ignorance depending on the state sent by Alice; (b) Bob can freely choose his measurement basis by applying the appropriate unitary transformation $U_b$. If he has access to the classical information possessed by Alice concerning her choice, all ignorance is of the classical kind. SPS: Single-Photon Source; P: Polarizer; PBS: Polarizing Beamsplitter; PC: Polarization Controller; D: Single-Photon Detector.}
\label{fig1}
\end{figure}

Let us consider the traditional scenario of quantum communication, where Alice sends a quantum state to Bob over a quantum channel. In the remainder of this discussion, for simplicity's sake, let us assume an ideal quantum channel, i.e., represented by the identity operator. The scenario is depicted in Fig. \ref{fig1}. Alice can prepare any pure polarization state (note the presence of a polarizer) with a general unitary transformation $U_a$ and send it to Bob, who can also apply any unitary transformation $U_b$ of his choice. The goal of this experiment is the same as in the previous section: Bob needs to guess which of his two detectors, $\text{D}_0$ or $\text{D}_1$, will fire, before the photon sent by Alice is effectively measured. Alice, in her turn, generates one out of four possible polarization states: $|H\rangle$, $|V\rangle$, $|+45^{\circ}\rangle = (|H\rangle + |V\rangle)/\sqrt{2}$ and $|-45^{\circ}\rangle = (|H\rangle - |V\rangle)/\sqrt{2}$.

Suppose, initially, that $U_b$ is a fixed transformation, which means that Bob always measures the polarization state of the photons he receives from Alice in the same basis, say, the horizontal-vertical (H-V) basis. Suppose Alice is also restricted to send $|H\rangle$ or $|V\rangle$ only, with equal probability (and this fact is known to Bob). The density operator representing the state as seen by Bob is then given by
\begin{equation}
\rho_1 = \frac{1}{2}|H\rangle \langle H| + \frac{1}{2}|V\rangle \langle V|
\label{eq1}
\end{equation}
which is equivalent to unpolarized light in the classical sense. In this case, Bob is bounded by classical ignorance: his odds of winning the game depend on how much (classical) information he has about Alice's choice. Representing Alice's choice by the random variable $A$ and Bob's guess by $B$, then the probability that Bob correctly guesses which detector fires is given by
\begin{equation}
P(\text{win}|\rho_1) = \frac{1+I(A,B)}{2}
\label{eq2}
\end{equation}
where $I(X,Y)$ is the mutual information of the random variables $X$ and $Y$. Clearly, if Bob's and Alice's choices are uncorrelated, Bob cannot guess which one of his detectors will fire with a probability greater than 50\%. But if Bob has full information on Alice's choices, he can always predict the measurement result with certainty.

However, if Alice sends the states $|+45^{\circ}\rangle$ and $|-45^{\circ}\rangle$ instead, then the situation changes completely. To see what happens, consider at first the expression for the new density operator, given by
\begin{equation}
\begin{array}{rll}
\rho_2 &= \frac{1}{2}|+45^{\circ}\rangle \langle +45^{\circ}| + \frac{1}{2}|-45^{\circ}\rangle \langle -45^{\circ}|\\
& = \frac{1}{2}|H\rangle \langle H| + \frac{1}{2}|V\rangle \langle V| 
\end{array}
\label{eq3}
\end{equation}
which is equal to $\rho_1$, as expected, as any incoherent superposition of equiprobable orthogonal pure states results in a completely mixed state. Even though the density operator is the same, Bob cannot predict any measurement result with certainty, even if he knows exactly which states Alice is sending him. For example, if Bob knows that Alice sent the polarization state $|+45^{\circ}\rangle$, detectors $\text{D}_0$ and $\text{D}_1$ will fire with probabilities $P(0) = |\langle +45^{\circ}|H\rangle|^2 = 1/2$ and $P(1) = |\langle +45^{\circ}|V\rangle|^2 = 1/2$, which results in
\begin{equation}
P(\text{win}|\rho_2) = \frac{1}{2}
\label{eq4}
\end{equation}
regardless of the value of $I(A,B)$. Bob's ignorance is, therefore, of the quantum kind.

Let us now assume that Bob has arbitrary control over his unitary transformation. He is allowed to preset any value he wants to $U_b$ prior to the detection (but not to keep it continuously changing during the measurement; this is a very important point that will be discussed in sec. IV). See how the situation has completely changed for Bob: if he has information on the states sent by Alice, he can previously adjust his unitary transformation such that his measurement basis matches the polarization state sent by Alice. Then, Bob will be able to predict with certainty his measurement result. In other words, Bob is now bound by classical ignorance.

Therefore, whenever Alice sends only pure states to Bob, and he is allowed to measure Alice's photons in any basis, Bob's ignorance is always of the classical kind. The density operator, in this case, represents his subjective ignorance on the (pure) state that describes the photon. Conversely, if Bob's classical ignorance is zero, he always knows the pure state which describes the photon's polarization. This is also true for all other degrees of freedom of the photon, as will be explained in the next section.

\section{\label{sec:4}The effect of polarization scramblers on single photons}

We have already discussed the controversial role of a polarization scrambler as a depolarizer. In order to make this point clear, let Alice's unitary operator be a function of time, $U_a(t)$, which randomly changes between the identity $I$ and the bit-flip operator $X$. If Alice's single photon source always produces photons in a fixed polarization state (e.g., vertical) and $U_a$ is always set before the photon is emitted, then Alice will produce an incoherent superposition of $|H\rangle$ and $|V\rangle$, just as in Eq.(\ref{eq1}). As we have already seen in sec. III, Alice is merely adding classical ignorance to her photons; the "unpolarized" aspect of Eq.(\ref{eq1}) merely represents Bob's ignorance on her choices for $U_a$. Therefore, the scrambler is not really depolarizing anything. This becomes even clear if, instead of single photons, Alice produces weak coherent pulses (phase-randomized coherent states): all photons that were emitted at the same time will be in the same pure state. Coherent measurements that measure two-photon polarization correlations, such as [ref], will immediately conclude that all pulses are polarized.

However, if Alice changes $U_a(t)$ during the coherence time of the photons, the situation changes. The same is true for Bob: we have shown in sec. III that the measurement of any pure state has zero quantum ignorance whenever the observer (Bob) is allowed to choose the measurement basis, which is represented by the unitary operator $U_b$.  However, it is important that this freedom in the choice of $U_b$ does not couple the polarization state to any other degrees of freedom of the photon being measured.

In order to make the above mentioned point clear, consider the scenario of Fig. 2(a), which is almost identical to the one considered before, but it has an unbalanced interferometer very much like the ones used in quantum key distribution systems that employ time-bin encoding \cite{Gisin_RMP}. The state of the photon just before the unitary transformation $U_a$ is given by
\begin{equation}
|\psi\rangle_A^{\text{in}} = \frac{1}{\sqrt{2}}(|s\rangle + e^{i\theta}|\ell\rangle)\otimes|H\rangle
\label{eq5}
\end{equation}
where the time modes $|s\rangle$ ("short") and $|\ell\rangle$ ("long") correspond to the early and late time bins, respectively. At the output of the polarization scrambler, the state of Eq. (\ref{eq5}) becomes
\begin{equation}
|\psi\rangle_A^{\text{out}} =  \frac{1}{\sqrt{2}}|s\rangle\otimes U_b(t_{s})|H\rangle +  \frac{1}{\sqrt{2}}e^{i\theta}|\ell\rangle\otimes U_b(t_{\ell})|H\rangle
\label{eqextra}
\end{equation}
Now consider that $U_a(t)$ is chosen such that in the time instant corresponding to the early mode we have $U_a(t_s) = I$ and in the late mode we have $U_a(t_{\ell}) = X$. The final state that is sent to Bob is, then, given by
\begin{equation}
|\psi\rangle_A^{\text{out}} =  \frac{1}{\sqrt{2}}|s\rangle\otimes|H\rangle +  \frac{1}{\sqrt{2}}e^{i\theta}|\ell\rangle\otimes|V\rangle
\label{eq6}
\end{equation}
which is a maximally entangled state between the polarization and time-bin degrees of freedom. There is a fundamental difference in this case concerning the (classical) information that Alice has on her state: even though she knows exactly which state she is sending, she does not (and cannot) know its polarization state, because it is undefined. So one might say that Bob's (and also Alice's) ignorance in a polarization measurement of the state represented by Eq. (\ref{eq6}) is of the quantum kind. Then we conclude that, in this case, the polarization scrambler is really depolarizing light.

\begin{figure}[htb]
\centerline{\includegraphics[width=8.5cm]{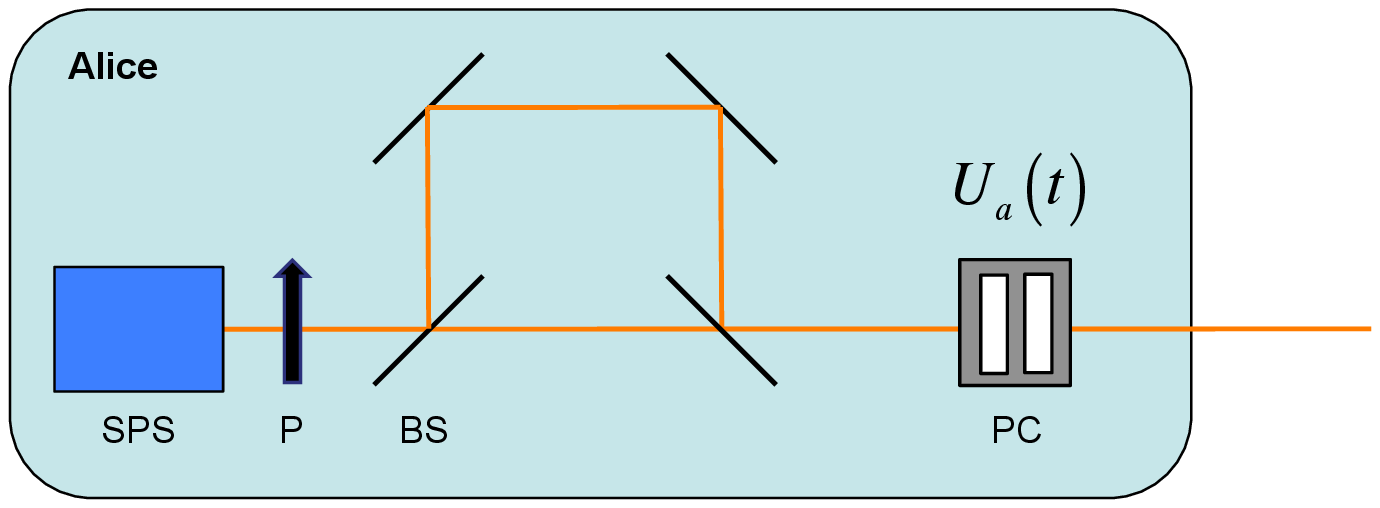}}
\centerline{\includegraphics[width=8.5cm]{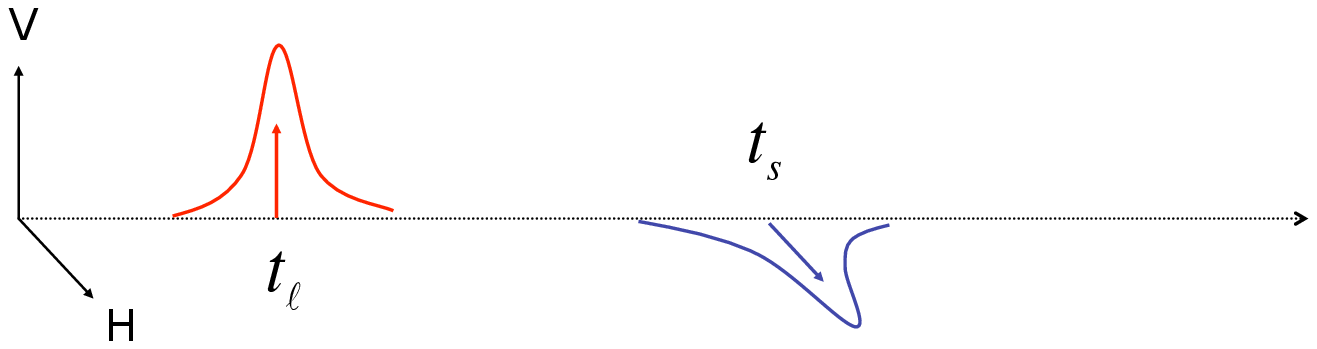}}
\caption{Alice sends polarization-time bin entangled photons to Bob. Up: Simplified scheme. The time delay introduced by the unbalanced interferometer is greater than the coherence time of the photons. Down: Output state generated by a time-varying unitary transformation such that $U_a(t_{s}) = I$ and $U_a(t_{\ell})=X$.}
\end{figure}

Bob, however, could still predict with certainty which detector will fire. If he knows which state was sent by Alice (i.e., his classical ignorance is zero), he could devise a very simple scheme that always results in a detection at $D_0$ (or $D_1$): in the expected arrival time of the early time bin, Bob selects $U_b = I$, whereas for the late time bin he chooses $U_b = X$, exactly as Alice has done. The effect of this operation is to undo the transformation introduced by $U_a$ such that the state $|\psi\rangle_A^{\text{out}}$ becomes again $|\psi\rangle_A^{\text{in}}$ (recall that $X^2 = I$). This state, in its turn, is fully polarized in the horizontal direction, so all detections would happen in $D_0$ (a similar scheme interchanging $I$ and $X$ would forward all photons to $D_1$). Bob would not, however, be able to determine the time of arrival of the photons.

The example above illustrates a very important point: if Bob has access to all degrees of freedom of the photon sent by Alice, his ignorance is always classical. This happens because Bob still sees the state sent by Alice as a pure state - the only difference is in the dimension of the Hilbert space, which has increased. Quantum ignorance only arises when Bob \textit{cannot} access all degrees of freedom. In particular, in the case of polarization measurement, Bob is only interested in measuring the polarization degree of freedom. The measurement realized with a time-varying $U_b$ is \textit{not} a standard polarization measurement: what Bob actually measured was a combination of two degrees of freedom, polarization and time.
 
In order to perform a pure polarization measurement, thus, Bob cannot be allowed to access higher dimensions of the Hilbert space describing the global state $|\psi\rangle_A^{\text{out}}$. In other words, he must restrict his unitary transformation to the form $U_b = U_{\text{pol}}\otimes I$, where $U_{\text{pol}}$ is a unitary transformation restricted to the Hilbert space representing the polarization state and $I$ is the identity on all other degrees of freedom. In the example that Alice sends the state of Eq.(\ref{eq6}), Bob is not allowed, therefore, to choose a unitary transformation that varies in a time scale shorter than the time difference between the early and late modes. In this case, Bob is completely bound by quantum ignorance, as all the information he has is represented by the reduced density operator of Eq. (\ref{eq6}), which is a completely mixed state. This is what we would call an "unpolarized" photon. Partially polarized photons could be produced using a similar method, e.g., replacing Alice's unitary operator such that $0 < |\langle H|U_a(t_{\ell})|H\rangle| < 1$.

\section{\label{sec:5}A new proposal for the degree of polarization}

It must already be clear that the DOP of a single photon should be based on the quantum ignorance in the polarization state measurement of this photon whenever the observer has the freedom of performing any von Neumann measurement he likes. This remark can also be stated as follows: the DOP of a photon must depend on the \textit{minimum} quantum ignorance of the measurement, where the minimum is taken over all possible measurements (i.e., among all possible bases). So the next question that naturally arises is: how can we quantify quantum ignorance in this situation?

We have already seen that quantum ignorance cannot be a function of the density operator, as density operators also take into account classical ignorance. As discussed in sec. III, zero classical ignorance is equivalent to know the pure state which represents all degrees of freedom of the photon; let $|\Psi_{\text{tot}}\rangle$ represent this pure state. Let $p_{\theta,\phi}$ and $1-p_{{\theta,\phi}}$ be the probabilities of detection in each detector in a pure polarization state measurement of state $|\Psi_{\text{tot}}\rangle$ with respect to the basis defined by $U_b({\theta,\phi})$, where $\theta$ and $\phi$ are the spherical coordinates that represent the basis in Poincar\'e Sphere. We define the \textit{minimum quantum ignorance} of this measurement as

\begin{equation}
Q_{\text{min}} = 1 - \text{sup}_{\theta,\phi}|1-2p_{\theta,\phi}|
\label{eq7}
\end{equation}
where $\theta \in [0,\pi]$, $\phi \in [0,2\pi]$, and
\begin{equation}
p_{\theta,\phi} = \text{tr}\left[U_b(\theta,\phi)|H\rangle \langle H|U_b^{\dagger}(\theta,\phi)\rho_{\text{pol}}\right]
\label{eq9}
\end{equation}
where
\begin{equation}
\rho_{\text{pol}} = \text{tr}_{\text{DOFs}}\left(|\Psi_{\text{tot}}\rangle \langle\Psi_{\text{tot}}|\right)
\label{eq10}
\end{equation}
and the partial trace of Eq. (\ref{eq10}) is over all degrees of freedom but polarization. Eq. (\ref{eq7}) can be interpreted as follows: we perform all possible pure polarization measurements in state $\rho_{\text{pol}}$ and register the probabilities $p_{\theta,\phi}$ and $1-p_{\theta,\phi}$ of projecting $\rho_{\text{pol}}$ onto each eigenstate describing the measurement for all values of $(\theta,\phi)$; or, equivalently, a single measurement is performed such that one of the probabilities $p_{\theta,\phi}$ or $1-p_{\theta,\phi}$ is maximized (the correct choice of the measurement which results in this maximum value is a consequence of the observer's zero classical ignorance). The quantum ignorance corresponds to the complement of the maximum difference between $p_{\theta,\phi}$ and $1-p_{\theta,\phi}$; this can be interpreted as the maximum "visibility" of the measurement. For example, if $\Psi_{\text{tot}}$ is a product state of the form $|\Psi_{\text{pol}}\rangle\otimes|\Psi_{\text{DOFs}}\rangle$, there will be a certain measurement that will result in probabilities 1 and 0 of projection in each eigenstate; thus the quantum ignorance is zero in this case. If $\rho_{\text{pol}}$ is a maximally mixed state as a result of entanglement with other degrees of freedom, as in Eq. (\ref{eq6}), then all probabilities $p_{\theta,\phi}$ will be equal to $1/2$, and the supremum in Eq.(\ref{eq7}) will be zero; hence, the quantum ignorance reaches its maximum value of 1.

Finally, the degree of polarization can be directly related to the quantum ignorance according to the simple formula
\begin{equation}
DOP = 1-Q_{\text{min}}
\label{eq8}
\end{equation}
which obviously satisfies $0 \leq DOP \leq 1$. As already discussed, the DOP is maximum for states with minimum quantum ignorance, which corresponds to all pure states or incoherent superpositions of pure states. This is a major difference between the classical concept of DOP and the quantity defined in Eq.(\ref{eq8}): as the classical DOP is a \textit{collective} property of all photons in a beam, any incoherent mixture of photons in orthogonal polarization states will always result in DOP = 0 (unpolarized light). The individual DOP of a photon, on the other hand, cannot be zero unless the polarization state is entangled with other degrees of freedom.

It should be mentioned that the definition given by Eq. (\ref{eq8}) is of counterfactual nature. The probabilities $p_{\theta,\phi}$ cannot be calculated with a single realization of the quantum system being considered, as in practice only one measurement can actually happen. This is in conformity to the fact that we cannot know the DOP of a single photon when we only have one copy of it.

\section{\label{sec:6}Final remarks}

We have provided a new definition for the degree of polarization, that is suitable for characterizing single photons, based on the minimum amount of quantum ignorance in a polarization state measurement of a photon. An analogy with a quantum communication scenario, where Bob has to guess which of his two detectors will fire before the photon sent by Alice arrives at his office, was provided in order to show that different sets of states sent by Alice that are classically interpreted as unpolarized light (and hence represented by the same density operator) can describe two opposite situations, provided that Bob can freely choose his measurement basis.  

If Alice sends an incoherent mixture of pure states, Bob is bounded by classical ignorance: whenever he has information on the quantum states sent by Alice, he can improve his odds by appropriately adjusting his measurement basis; if Alice sends a state in which the polarization state is entangled with other degrees of freedom that Bob cannot access, on the other hand, Bob's ignorance is of the quantum kind: no matter the amount of classical information he has access to, his probability of correctly guessing the outcome of the measurement is always smaller than one. Both situations can be generated with a polarization scrambler, which means that it is able to depolarize light under certain conditions.

The proposed measure of degree of polarization, differently from the classical definition, does not rely on the description given by the density operator: it actually considers that the observer has access to all possible classical information on the state, just as if he had prepared it, which corresponds to a pure state in a higher-dimensional Hilbert space. It is, therefore, an objective quantity rather than a subjective one.

 \begin{acknowledgments}
The authors acknowledge financial support from CNPq and FAPERJ.
\end{acknowledgments}

% The \nocite command causes all entries in a bibliography to be printed out
% whether or not they are actually referenced in the text. This is appropriate
% for the sample file to show the different styles of references, but authors
% most likely will not want to use it.
%\nocite{*}

%\bibliography{apssamp}% Produces the bibliography via BibTeX.

\begin{thebibliography}{000}
\bibitem{Comment1}
In practice there is, in fact, a difference between the two scenarios: the first one corresponds to a classical (pseudo) random number generator (RNG), whereas the second one is equivalent to a quantum RNG.
\end{thebibliography}

\end{document}